
\magnification=1200
\baselineskip=20pt
\font\tipo=cmr10
\font\it=cmti10

\font\tit=cmbx10

\font\ref=cmti9

\hoffset=1.6truecm
\voffset=1.0truecm
\hsize=6truein
\vsize=8.5truein
\baselineskip12pt
\centerline{\tit THE CAYLEY-HAMILTON THEOREM FOR SUPERMATRICES}

\vskip3cm

\tipo
\baselineskip12pt
\centerline{L.F. Urrutia}
\centerline{Instituto de Ciencias Nucleares}
\centerline{Universidad Nacional Aut\'onoma de M\'exico}
\centerline{Circuito Exterior, C.U.}
\centerline{04510 M\'exico, D.F.}

\centerline{and}

\centerline{Centro de Estudios Cient\'\i ficos de Santiago}
\centerline{Casilla 16443, Santiago 9}
\centerline{Chile}

\

\centerline{and}

\

\centerline{N. Morales}
\centerline{Department of Mathematics}
\centerline{Universidad Aut\'onoma Metropolitana-Iztapalapa}
\centerline{Apartado Postal 55-534}
\centerline{09340 M\'exico, D.F.}


\vfill\eject

\vskip2cm
\baselineskip20pt

\centerline{\tit ABSTRACT}

\

Starting from the expression for the superdeterminant of
$ (xI-M)$, where
$M$  is an arbitrary supermatrix , we propose a definition for the
corresponding
characteristic polynomial and we prove that each supermatrix
satisfies its  characteristic equation.
Depending upon the factorization properties of the  basic polynomials
whose ratio defines  the above mentioned superdeterminant we are
able to construct polynomials of lower degree which are also shown to be
annihilated
by the supermatrix. Some particular cases and examples are discussed.

\vfill\eject

\tipo

\noindent
1.- INTRODUCTION

Given any $n \times n$ real matrix $M$, its characteristic
polynomial is defined by $P(x) = det (xI - M)$, where $I$ denotes
the $n \times n$ identity matrix and $x$ is a real variable. In
general $P(x) = x^n + \sum_{k=0}^{n-1} c_k x^k$ is a monic
polynomial of degree $n$. The Cayley-Hamilton theorem asserts that
$P(x = M) = 0$. That is to say, if we substitute in $P(x)$ the
real variable $x$ by the matrix $M$ in all the powers $x^k (k \not=
0)$, and set $x^0 = I$, we obtain the matrix zero as the result.
This is a powerful theorem in the sense that it produces $n^2$ null
identities among the matrix elements. The coefficients $c_k (k
\not= 0)$ can be written in terms of $Tr(M), Tr(M^2),\cdots ,
Tr(M^{n-1})$ together with their powers and $c_0 = det (M)$.
This theorem has recently found interesting applications
in 2+1 dimensional Chern-Simons  (CS) theories [1]. Pure CS
theories are of topological nature and the fundamental degrees
of  freedom are the traces of group elements constructed
as the holonomies ( or Wilson lines, or integrated  connections)
of the gauge connection around oriented closed curves on the
manifold. The observables are the expectation values of the Wilson lines which
turned out to be realized as the various knot polynomials
known to  mathematicians [2]. Since CS theories are also exactly
soluble and possess a finite number of degrees of freedom [3], another
aspect of interest is the reduction of the initially infinite-dimensional phase
space to the subspace of the true degrees of
freedom .The Cayley-Hamilton theorem has played an important
role in the construction of the so called skein  relations [4],
which are relevant to the calculation of expectation values, and
also in the process of reduction of the phase space. To illustrate
the basic ideas related to this last point let us consider
the simple case of
two matrices $M_1$ and $M_2$ which
belong to $SL(2,R)$. In this case the characteristic polinomial
is $P(x) = x^2 - Tr(M_1) x + 1$ and we have the
Cayley-Hamilton matrix identity

$$(M_1)^2 - Tr(M_1) M_1 + I = 0. \eqno(1.1)$$



\noindent
By multiplying Eq. (1.1) by $M_2 M_1^{-1}$ and tracing we obtain the
following relation among the traces

$$Tr(M_2 M_1^{-1}) + Tr(M_1 M_2) = Tr(M_1) Tr(M_2).
\eqno(1.2)$$

\noindent
We recall also that for any $SL(2,R)$ matrix we have $Tr(M) =
Tr(M^{-1})$. The expression (1.2) finds a very useful application
in the discussion of the reduced phase space of the de Sitter
gravity in 2 + 1 dimensions, which is equivalent to the
Chern-Simons theory of the group $SO(2,2)$ [3]. This theory can be
more easily described in terms of two copies of the group
$SL(2,R)$, which is the spinorial group of $SO(2,2)$. The gauge
invariant degrees of freedom associated to one genus of an
arbitrary genus $g$ two-dimensional surface turn out to be traces
of any product of powers of two $SL(2,R)$ matrices $M_1$ and
$M_2$, which correspond to the holonomies (or integrated
connections) of the two homotopically distinct trajectories on
one genus. Nevertheless, because Chern-Simons theories have a
finite number of degrees of freedom, one should be able to reduce
this infinite set of traces to a finite one. This task can in
fact be accomplished by virtue of the relation (1.2). In other
words, the trace $Tr(M_1{}^{p_1} M_2{}^{q_1} M_1{}^{p_2}
M_2{}^{q_2} \cdots M_1{}^{p_n} M_2{}^{q_n} \cdots )$, for any
$p_i, q_i$ in $\cal{Z}$, can be shown to be reducible and to be expressed
as a function of three traces only: $Tr(M_1), Tr(M_2)$ and
$Tr(M_1 M_2)$ [5]. A simple example of such reduction is to consider
$Tr(M^2_1 M_2)$ for example. Here we apply the relation (1.2)
with $M_1 \to M_1$ and $M_2 \to M_1 M_2$ obtaining

$$Tr(M_1^2 M_2) = Tr(M_1) Tr(M_1 M_2) - Tr(M_2). \eqno(1.3)$$

\noindent
A similar reduction can be performed in the case of 2 + 1 super
de Sitter gravity, which is the Chern-Simons theory of the
supergroup $Osp(2\vert 1, C\hskip-7pt I \ )$ [6]. The novelty here is that one
is
dealing with supermatrices instead of ordinary matrices. In the
particular case considered, a Cayley-Hamilton identity for the
supermatrices was obtained in an heuristical way and a relation
analogous to (1.2) was derived. This allowed to carry out the
reduction of the infinite dimensional phase space, this time in
terms of five complex supertraces [7]. We observe that the  non-linear
constraints among the traces that need to be solved in order to accomplish the
reduction of the phase space, of which Eq.(1.2) is an example, are ussually
obtained using the so called Mandelstam identities [8].
The  discussion of the relation among these two alternatives together with the
consequences of using
such identities in the case of supermatrices is left for future
work.

In this paper we discuss the general construction of
Cayley-Hamilton type identities for supermatrices. This is an
interesting problem in its own, besides the possible applications
in the study of the reduced space in Chern-Simons theories
defined over a supergroup. In Section 2 we introduce our notation
together with a number of results which will be useful for
our purposes. In this section we also introduce a definition of
the characteristic and null polynomials for supermatrices starting from the
corresponding superdeterminant. In Section 3 we prove the
Cayley-Hamilton theorem for the polynomials
previously defined, by introducing the analogous of the adjoint
for supermatrices. The main results contained in Sections 2 and 3 have been
already reported as a Letter in Ref. [9]. They are included here to make this
paper selfcontained and also to allow for a more detailed presentation. Section
4 contains a
discussion of some interesting cases together with many specific
examples. Finally, in Section 5 we give a short summary of this work
emphasizing those points, that in our opinion, require further developement.
There are also two Appendices where some useful results
are collected.

\vskip 1.4pc
\noindent
2.- THE CHARACTERISTIC AND NULL POLYNOMIALS FOR SUPERMATRICES

A $(p+q) \times (p+q)$ supermatrix is a block matrix of the form

$$M= \left( \matrix{A & B\cr
C & D\cr}
\right), \eqno(2.1)$$

\noindent
where $A, B, C$ and $D$ are $p\times p, p \times q, q\times p,
q\times q$ matrices respectively. The distinguishing feature with
respect to an ordinary matrix is that the matrix elements
$M_{RS} \ R= (i, \alpha), \ S= (j,\beta)$ are elements of a Grassmann
algebra with the property that $A_{ij} \ (i,j = 1, \cdots p)$ and
$D_{\alpha\beta} \ (\alpha,\beta = 1, \cdots q)$ are even
elements, while $B_{i\alpha}$ and $C_{\beta j}$ are odd elements
of such algebra. In particular this means that such numbers satisfy

$$\eqalign{B_{i\alpha} B_{j\beta} = -&B_{j\beta} B_{i\alpha}, \  C_{\alpha i}
C_{\beta j} = - C_{\beta j} C_{\alpha i} \cr
&B_{i\alpha} C_{\beta j} = -C_{\beta j} B_{i\alpha}, \cr}\eqno(2.2)$$

\noindent
while $A_{ij}$ and  $D_{\alpha\beta}$
commute with everything.

Let us recall that the ordinary matrix addition and the ordinary
matrix product of two supermatrices is again a supermatrix.
Nevertheless, such concepts as the trace and the determinant need
to be redefined, because of the odd component piece of the
supermatrix.

The basic invariant under similarity transformations for
supermatrices is the supertrace, defined by

$$Str(M) = Tr(A) - Tr(D), \eqno(2.3)$$

\noindent
where the trace over the even matrices is the standard one.  An
important property of the above definition is the cyclic identity
$Str(M_1M_2) = Str (M_2M_1)$, for arbitrary supermatrices, which
is just a consequence of the relative minus sign in (2.3). The
generalization of the determinant, called the superdeterminant,
is obtained from (2.3) by defining

$$\delta ln (Sdet M) = Str (M^{-1}  \delta M), \eqno(2.4)$$

\noindent
with apropriate boundary conditions. In this compact notation
we are sumarizing the $(p+q)^2$ relations which give the partial
derivatives of the function $ln(Sdet M)$ with respect to the
entries $M_{RS}$ of the supermatrix, in terms of the elements of
the inverse supermatrix $M^{-1}$. For example, ${\partial ln(Sdet
M)\over \partial M_{ij}} = (M^{-1})_{ji}$ for the even indices
$i,j$. These first order partial differential equations are
subsequently integrated under the boundary conditions $Sdet I =
1$, where $I$ is the unit supermatrix, to produce the following
equivalent two forms of calculating the superdeterminant [10]

$$Sdet (M) = {det(A-B D^{-1} C)\over det D} = {det A \over
det(D-C A^{-1} B)}. \eqno(2.5)$$

\noindent
Here all the matrices involved now are even in the Grassmann
algebra and $det$ has its usual meaning. The superdeterminant
inherits the basic property
$Sdet(M_1 \break  M_2) = Sdet (M_2 M_1)$ and requires $det D \not= 0 $
and $det A \not= 0$ in order to be defined. An explicit
demonstration of the equality of the two alternatives ways of
calculating $Sdet (M)$ is given in the Appendix $A$.

In order to proceed we introduce $a(x) = det (xI-A)$ and $d(x) =
det (xI-D)$, which are the characteristic polynomials of the even
matrices $A$ and $D$.

Let us consider now the function $h(x) = Sdet (xI-M)$ which could
be naively taken as the analogous of the standard characteristic polynomial.
Nevertheless, this function is in fact the ratio of two
polynomials.

$$h(x) = {\tilde F(x)\over \tilde G(x)} = {F(x) \over G (x)},
\eqno(2.6)$$

\noindent
each form arising from the two alternatives (2.5) of calculating
the superdeterminant. The explicit expressions for the numerators
and denominators are

$$\tilde F(x) = det (d(x) (xI-A) - B adj(xI-D) C), \ \ \tilde
G(x) = (d(x))^{p+1}, \eqno(2.7a)$$

$$F(x) = (a(x))^{q+1}, \ \ G(x) = det (a(x) (xI-D) - C adj
(xI-A)B), \eqno(2.7b)$$

\noindent
where we have used the basic relation $(xI-F)^{-1} = [det(xI-F)]^{-1} adj
(xI-F)$ valid for any even matrix $F$. Notice that $\tilde F$ is
expressed in terms of the determinant of a $p \times p$ even
matrix, while $G(x)$ is the determinant of a $q \times q$ even
matrix.

In order to motivate the basic idea of our definition for the
characteristic polynomial of a supermatrix let us consider the
simple case of a block-diagonal supermatrix $M \ (i.e. \ B = 0, C =
0)$. Here $h(x) = a(x)/d(x)$ and clearly the characteristic
polynomial is $P(x) = a(x) d(x)$. In fact we have

$$P(M) = \left( \matrix{ a(A) & 0\cr
0 & a(D) \cr}\right) \
\left(\matrix{ d(A) & 0\cr
0 & d(D)\cr} \right) \equiv 0 \eqno(2.8)$$

\noindent
because $a(A) = 0, d(D) = 0$. In the case where $a(x)$ and
$d(x)$ have a common factor $f(x)$

$$a(x) = f(x) a_1(x) , \ d(x) = f(x) d_1 (x), \eqno(2.9)$$

\noindent
the answer is given by $P(x) = f(x) a_1 (x) d_1(x),$
which is a polynomial of lower degree than the product $a(x)
d(x)$.

In terms of the superdeterminant (2.6) the proposed definition of the
characteristic polynomial $ {\cal P} \left(x \right)$
for an arbitrary supermatrix is

$${\cal P}(x) = \tilde F (x) G(x) = F(x) \tilde G(x) = a(x)^{q+1}
d(x)^{p+1}. \eqno(2.10)$$
For notational simplicity we will not necessarily write
explicitly the
x-dependence on many of the polynomials considered
in the sequel.

Nevertheless, and motivated by the work of Ref. [11], we have
realized that there are some cases in which we can construct null
polynomials of lower degree according to the factorization
properties of the basic polynomials $\tilde F, \tilde G, F, G$.
At this point it is important to observe that we do not have a
unique factorization theorem for polynomials defined over a
Grassmann algebra. This can be seen,  for example,  from the identity $x^2 =
(x+\alpha)(x-\alpha)$, where $\alpha$ is an even Grassmann with
$\alpha^2 = 0$. The construction of such null polynomials of lower degree
starts
from finding the divisors of maximun degree of the pairs $\tilde
F, \tilde G, (F,G)$ which we denote by $R(S)$ respectively. This
means that one is able to write.

$$\eqalign{\tilde F = R \tilde f, \ \  \tilde G = R \tilde g, \cr
F = Sf, \ \  G = Sg,\cr}\eqno(2.11)$$

\noindent
where all polynomials are monic and also $\tilde f, \tilde g, f,
g$ are of least degree by construction. They must satisfy

$$\tilde f/\tilde g = f/g  \eqno(2.12)$$

\noindent
because of the Eq. (2.6) and the expressions in (2.11) might be
not unique. Let us observe that in the case of polynomials over
the complex numbers Eq. (2.12) would imply at most $\tilde f =
\lambda f, \tilde g = \lambda g$ with $\lambda$ being a constant.
Since we are considering polynomials over a Grassmann algebra
this is not necessarily true as can be seen again in the above
mentioned identify $x/(x- \alpha) = (x+\alpha)/x$, which we have
rewritten in a convenient way. For each family of possible
factorizations written in Eq. (2.11) we define a null polynomial
$P(x)$ as

$$P(x) = \tilde f(x) g (x) = f(x) \tilde g (x),  \eqno(2.13)$$

\noindent
which is clearly of less degree than ${\cal P}(x)$. The
characteristic polynomial  ${\cal P}(x)$ is just a particular case
of these null polynomials when $R=S=1$ so that we will
concentrate mostly on the definition (2.13).

\vskip 1.4pc
\noindent
3.- THE CAYLEY-HAMILTON THEOREM FOR SUPERMATRICES

Part of our strategy to prove such theorem for the
polynomial introduced in Eq. (2.13) is based in one of the
standard methods to prove the Cayley-Hamilton theorem for
ordinary matrices [12]. We briefly recall such procedure and
emphasize that it is independent of the matrix considered been a
standard matrix or a supermatrix.

The basic point is the following: if for the supermatrix $(xI-M)$,
with $M$ being an $n\times n$ supermatrix independent  of
$x$, there exists a polynomial supermatrix $N(x) = N_0^{m-1}
+ N_1 x^{m-2} + .. + N_{m-1} x^0$, (where each $N_k \ (k = 0,
\cdots , m-1)$ is an $n\times n$ supermatrix independent of $x$) such
that

$$(xI-M) N(x) = P(x) I, \eqno(3.1)$$

\noindent
where $P(x) = p_0 x^n + p_1 x^{n-1} + \cdots + p_n x^0$ is a
numerical polynomial over the Grassman algebra, then $P(M) = p_0 M^n + p_1
M^{n -1} + \cdots + p_n I \equiv 0$. The proof follows by
comparing the independent powers of $x$ in Eq. (3.1). The
equality of the highest power in $x$ on both sides implies $m =
n$, together with $N_0 = p_0 I$. This leads
to the following set of supermatrix relations for the remaining powers

$$\eqalign{ N_1 - p_0  M &= p_1I,\cr
N_2 - MN_1 &= p_2 I.\cr
N_{n-1} - MN_{n-2} &= p_{n-1} I,\cr
-MN_{n-1} &= p_n I.\cr}\eqno(3.2)$$

\noindent
By multiplying on the left the $k-th$ equation by $M^{n-k}$
and adding up all the relations we can verify that all the terms
involving $N_k$ cancel out in the LHS leaving only the
term $-p_0 M^n$. The RHS is just $P(M)- p_0 M^n$ so that we have the
result $P(M) = 0$ as required. In the standard case the matrix
$N(x)$ is just given by $N(x) = adj (xI-M) = det(xI-M)
(xI-M)^{-1}$, and $P(x) = det (xI-M)$.

In the case of a supermatrix we do not have an obvious
generalization either of the polynomial matrix $adj (xI-M)$ or
of $det (xI-M)$. Nevertheless, following the analogy as close as
possible we define

$$N(x) = P(x) (xI-M)^{-1} \eqno(3.3)$$

\noindent
where $P(x)$ is the polynomial (2.13)
introduced in the previous section. The challenge now is to prove
that $N(x)$, which trivially satisfies the Eq. (3.1), is indeed a
polynomial matrix. In this way we would have proved that $P(M) =
0$, according to our previous argument based upon relations (3.2).

In the first place we calculate $(xI-M)^{-1}$ in block form, with
the result

$$(xI-M)^{-1}_{11} = ((xI-A)-B (xI-D)^{-1} C)^{-1}, \eqno(3.4a)$$

$$(xI-M)^{-1}_{12} = (xI-A)^{-1}B ((xI-D) - C(xI-A)^{-1} B)^{-1},
\eqno(3.4b)$$

$$(xI-M)^{-1}_{21} = (xI-D)^{-1}C ((xI-A)-B (xI-D)^{-1}C)^{-1},
\eqno(3.4c)$$

$$(xI-M)^{-1}_{22} = ((xI-D)-C (xI-A)^{-1} B)^{-1}, \eqno(3.4d)$$

\noindent
where the subindices 11, 12, 21 and 22 denote the corresponding
$p\times p, p\times q, q\times p$, and $q\times q$ blocks
respectively. Notice that the conditions for the existence of
$(xI-M)^{-1}$ are the  same as those for the existence of $Sdet
(xI-M)$ and they are $det(xI-A) \not= 0$ and $det(xI-D) \not=
0$. Since $x$ is a generic Grassmann variable we will assume that
this is always the case. By virtue of these assumptions the term
$((xI-A) - B(xI-D)^{-1} C)^{-1}$, for example, can always be calculated as
$(I-(xI-A)^{-1}B (xI-D^{-1}) C)^{-1} (xI-A)^{-1}$. The factor on the
left can be thought as a series expansion of the form $1/(1-z) = 1 +z+z^2+
\cdots ,$ with $z= (xI-A)^{-1}B
(xI-D)^{-1} C$. Moreover, the series will stop at some power
because $z$ is a matrix with body zero and thus it is nilpotent.

Before going to the specific
case for $P(x)$ considered in Section 2 it will prove most
convenient to rewrite the expressions (3.4) in terms of the
functions $F, G, \tilde F $ and $\tilde G$, together with $a$ and
$d$. To begin with, we express the inverse of all even matrices
in (3.4) in terms of their adjoint and the corresponding
determinant. We obtain

$$(xI-M)^{-1}_{11}= {d\over \tilde F} adj ( (xI-A)d - Badj
(xI-D) C), \eqno(3.5a)$$

$$(xI-M)^{-1}_{12}= {1\over G} adj (xI-A)B adj
((xI-D)a - C adj (xI-A) B),\eqno(3.5b)$$

$$(xI-M)^{-1}_{21}= {1\over \tilde F} adj  (xI-D) C adj
((xI-A)d - B adj (xI-D) C),\eqno(3.5c)$$

$$(xI-M)^{-1}_{22}= {a\over G} adj ( (xI-D)a - Cadj
(xI-A) B),\eqno(3.5d)$$

\noindent
recalling the definitions of $\tilde F$ and $G$ in Eq. (2.7). Next
we proceed by expressing the above matrix elements as derivatives
of the even functions $\tilde F$ and $G$ with respect to the
generic supermatrix elements $A_{ij}, B_{i\alpha}, C_{\alpha i},
D_{\alpha\beta}$. To this end we use the basic property

$$\delta det Q = Tr (adj Q \delta Q)\eqno(3.6)$$

\noindent
valid for any even matrix $Q$. The meaning of this compact
notation has been already explained in the paragraph after Eq.
(2.4) and, mutatis mutandis, applies also to even matrices. Let
us consider the variation of
$\tilde F$ with respect to $B_{i\alpha}$, keeping all the other
entries constant. We obtain

$$\delta\tilde F = -[adj ((xI-A) d- B
adj (xI-D) C)]_{ij} \delta B_{j\alpha} [adj (xI-D) C]_{\alpha
i}, \eqno(3.7)$$

\noindent
which means that

$${\partial\tilde F\over \partial B_{j\alpha}} = - [adj (xI-D) C
adj ((xI-A) d-B adj (xI-D) C)]_{\alpha j}. \eqno(3.8)$$

\noindent
Here we are taking the derivative with respect to an odd
Grassmann number from the left in the sense that $\delta\tilde F
\equiv \delta B_{j\alpha} {\partial\tilde F\over \partial
B_{j\alpha}}$.
 We have then succeeded in writing
the matrix elements of $(xI-M)^{-1}_{21}$ as

$$(xI-M)^{-1}_{\alpha j} = - {1\over \tilde F} {\partial \tilde
F\over \partial B_{j\alpha}}. \eqno(3.9)$$

\noindent
In a completely analogous way we obtain

$$(xI-M)^{-1}_{i\alpha} = - {1\over G} {\partial G\over
\partial C_{\alpha i}}. \eqno(3.10)$$

\noindent
Now let us rewrite the block-diagonal terms. To this end it is
convenient to look for variations with respect to $A_{ij}$
and $D_{\alpha\beta}$. The change of $\tilde F$ with respect
to $A_{ij}$, keeping all the other entries of M constants is  given by

$$\delta\tilde F = - d \left[ adj ((xI-A)d - B adj
(xI-D)C\right]_{ij} \ \delta A_{ji}, \eqno(3.11)$$

\noindent
which can be written as

$${\partial\tilde F\over \partial A_{ji}} = - d
\left[ adj ((xI-A)d - B adj (xI-D) C)\right]_{ij}. \eqno(3.12)$$

\noindent
In other words we have shown that

$$(xI-M)^{-1}_{ij} = - {1\over \tilde F}{\partial \tilde F\over
\partial A_{ji}}. \eqno(3.13)$$

\noindent
In a completely analogous way, the block-diagonal term 22 can be
expressed as

$$(xI-M)^{-1}_{\alpha\beta} = - {1\over G} {\partial G\over
\partial D_{\beta\alpha}}.\eqno(3.14)$$

Now we come to the last step of our argument which consists in using
the polynomial (2.13) defined in the previous section
together with the factorization properties (2.11) to
prove that $N(x) = P(x) (xI-M)^{-1}$ is
a polynomial matrix.

Let us consider the block-elements 11 and 21 of
$P(x) (xI-M)^{-1}$ to begin with. According to the expression
(3.13) together with (2.11), the first one can be written as

$$N_{ij} = - g {\partial \tilde f\over \partial A_{ji}} -
{g \tilde f \over R} {\partial R\over \partial A_{ji}}.
\eqno(3.15)$$

\noindent
The first term of the LHS is clearly of polynomial character. In
order to transform the second term we make use of the property

$${\partial ln \tilde G\over \partial A_{ji}} = 0 = {\partial ln
R\over \partial A_{ji}} + {\partial ln \tilde g \over \partial
A_{ji}}, \eqno(3.16)$$

\noindent
which follows from the factorization $\tilde G = R \tilde g$,
together with the fact that $\tilde G$ is just a function of
$D_{\alpha\beta}$, according to Eq. (2.7b). In this way, and using  also the
Eq.(2.12),  we obtain

$$N_{ij} = f {\partial \tilde g\over \partial A_{ji}} -
g {\partial \tilde f\over \partial A_{ji}}, \eqno(3.17)$$

\noindent
which leads to the conclusion that the block-matrix $N_{ij}$ is
indeed polynomial. The proof for $N_{\alpha i}$ runs along the
same lines, except that now the derivatives are taken with
respect to $B_{i\alpha}$ and that we have to use ${\partial ln
\tilde G\over \partial B_{i\alpha}}= 0$, instead of Eq. (3.16).

The remaining terms $N_{i\alpha}$ and $N_{\alpha\beta}$ can be
dealt with in analogous manner by considering the derivatives of
$ G =S g$ with respect to $C_{\alpha i}$ and
$D_{\beta\alpha}$, and by replacing the condition (3.16) by
${\partial ln F\over \partial C_{\alpha i}} = 0 $ and ${\partial
ln F\over \partial D_{\beta\alpha}} = 0 $ respectively. The
results are again of the form (3.17), the only difference been the
variables with respect to which the derivatives are taken.

To summarize, we have proved that $N(x) \equiv P(x) (xI-M)^{-1}$
is a polynomial matrix, for $P(x)$ defined in Eq. (2.13). Then
it follows immediately that $(xI-M) N(x) = P(x) I$ and from the
demonstration at the beginning of Section 3 we obtain the desired
result $P(M) = 0$. The same conclusion holds for the
characteristic polynomial ${\cal P}(x)$ defined in Eq. (2.10). This
concludes the proof of the Cayley-Hamilton
theorem for supermatrices. Let us remark that the supermatrix
$N(x)$ can be considered to be the generalization of $adj(xI-M)$ to
the case of supermatrices.

\vskip 1.4pc
\noindent
4.- PARTICULAR CASES AND SPECIFIC EXAMPLES.

In this section we consider some distinguished cases and some
particular examples of null polynomials of minimum degree for
supermatrices, constructed according to the definitions given in
Section 2. Our general procedure for constructing such null
polynomials is based in the factorization properties of the
 polynomials $\tilde F, \tilde G, F, G$ introduced in
Section 2. The work of Ref. [11]  shows that these factorization
properties are closely related to those of the characteristic
polynomials $a(x)$ and $d(x)$ corresponding to the even blocks of
the supermatrix. At this point we emphasize that when dealing
with polynomials over a Grassmann algebra, the existence of a
maximum common divisor of two polynomials is not in one to one
correspondence with the fact that these polynomials are not
coprime. In fact, we will exhibit a simple example of two
polynomials which are not coprime and nevertheless do not have a
common factor. In this section we will shift the emphasis to the
factorization properties of $a(x)$ and $d(x)$ and we will
consider three cases: (1) the polynomials $a$ and $d$ are
coprime, (2) the polynomials $a$ and $d$ are not coprime but do
not have a common factor and finally (3) both polynomials are not
coprime and have a maximum common divisor.

The first case has been thoroughly discussed in Ref. [11]  in
relation with the factorization properties of $h(x)$. Theorem
(3.9) of this reference proves that the
superdeterminant $h$ can be written in the unique irreducible
form

$$h(x) = (a+r)/(d+t)  \eqno(4.1)$$

\noindent
where $r$ and $t$ are even polinomials with body zero which have
the property $deg (r) < deg (a)$ and $deg(t) < deg(d)$.
The two basic steps that lead to the expression (4.1) are, in
the first place, the possibility of writting

$$\tilde F = a d^p + u, \ G = a^q d + v,
\eqno(4.2)$$

\noindent
together with the factorization

$$\tilde F = (a + r) (d^p + t^\prime), \eqno(4.3a)$$

$$G =  (d+ t) (a^q + r^\prime), \eqno(4.3b)$$

\noindent
where all polynomials $u, v, r, r^\prime, t, t^\prime$ have body
zero and $deg (u) < p(q+1), deg(v) < q(p+1), deg(t^\prime) <
pq, deg(r) < p, deg (r^\prime) < pq, deg (t) < q$. The
expressions (4.2) are just the expansions
of (2.7a) and (2.7d) in terms of powers of the odd generators
$B_{i\alpha}$ and $C_{\beta j}$, while the expressions (4.3)  and (4.4) are
a consequence of corollary (3.8) of Ref. [11], which we have
included in our Appendix $B$ for further use. The second
step arises form comparing the two ways (2.6) of writting
$h(x)$ and using the factorization lemma (3.4) of Ref. [11],
also included in our Appendix $B$. In this way one obtains that

$$F = a^{q+1} = (a+r) (a^q + r^\prime), \eqno(4.4a)$$

$$\tilde G = d^{p+1} = (d+t) (d^p + t^\prime). \eqno(4.4b)$$

\noindent
The use of (4.3) and (4.4) in any expression (2.6) leads
directly to the form (4.1) for $h(x)$. Our final conclusion
for the case where $a(x)$ and $d(x)$ are coprime is
the following expression for the null polynomial of minimum degree

$$P(x) \equiv (a + r) (d + t),\eqno(4.5)$$

\noindent
according to the ideas developped in Section 2 and 3. Besides
giving all
these existence theorems, we can find in Ref. [11]  what they
call a modified Euclidean algorithm, which in fact allows to
explicitly perform the reduction in Eq. (2.6) thus obtaining the
irreducible expressions appearing in Eq. (4.1).

The simplest example of this case correspond to the $(1+1)\times (1+1)$
supermatrix

$$M = \left( \matrix{ p & \alpha \cr
\beta & q \cr}\right), \eqno(4.6)$$

\noindent
with $\bar p \not= \bar q$ in such a way that $a=x-p$ and $d
= x-q$ be coprime polynomials acoording to lemma (3.3) of Ref.[11]. The bar
over a number denotes its body.  Here we have

$$\eqalign{\tilde F = (x-q) (x-p)-\alpha\beta, \ \ \ \  \tilde G =
(x-q)^2 \cr
F = (x-p)^2 , \ \ \ \  G = (x-q) (x-p) + \alpha\beta.\cr} \eqno(4.7)$$

The modified Euclidean algorithm of Ref. [11] applied for each
pair $\tilde F, \tilde G \ \ \ (F, G)$ leads to the following factorizations

$$\eqalign{ \tilde F &= (x-p + {\alpha\beta \over q-p}) (x-q -
{\alpha\beta\over q-p}),\cr
\tilde G &= (x-q + {\alpha\beta\over q-p})(x-q -
{\alpha\beta\over q-p}),\cr
F &= (x-p + {\alpha\beta \over q-p}) ( x-p - {\alpha\beta \over
q-p}),\cr
G &= (x-q + {\alpha\beta \over q-p})( x - p - {\alpha\beta \over
q-p}), \cr}\eqno(4.8)$$

\noindent
which allow the identifications $r = -t^\prime = t =
-r^\prime = {\alpha\beta \over q - p}$, since these polynomials
are of degree zero in this case. Here we have $R = (x-q -
{\alpha \beta \over q-p}), \ S = (x-p - {\alpha \beta \over q -
p})$ together with $\tilde f = f = x-p + {\alpha\beta \over q -
p}$ and $ g = \tilde g = (x-q + {\alpha\beta \over q - p})$ in
the notation of Section 2. The null polynomial of minimum degree
is then [9, 13]

$$P(x) = fg = x^2- x(p+q-{2\alpha\beta \over q-p}) + pq -
{(q+p)\over (q-p)} \alpha\beta, \eqno(4.9)$$

\noindent
where  we  can verify by direct substitution that $P(M) \equiv 0$.

Another example of this kind corresponds to the case of
supermatrices belonging to the supergroup $Osp(1\vert 2; C \hskip
-7pt / )$ which are relevant to the discussion of the reduced  phase space in
super de Sitter
Gravity in 2+1 dimensions  [6, 7 ]. Here we consider $(2+1)\times (2+1)$
supermatrices so that the factorization properties involved are a
particular case of the example developped in Ref. [11].

The supermatrices $M$ belonging to $Osp(1\vert 2; C \hskip -7pt /
)$ are such that

$$M^T H M = H; \ \  H = \left( \matrix{ 0 & 1 & 0 \cr
-1 & 0 & 0\cr
0 & 0 & 1 \cr}\right) \eqno(4.10)$$

\noindent
where $T$ denotes the supertransposed and $H$ is the
orthosimplectic supermatrix. The above supermatrices can be
parametrized in the following way,

$$ M = \left( \matrix{ A & \xi \cr
\eta^t & a \cr} \right), \ \xi =  \left( \matrix{ x_1 \cr
x_2 \cr}\right), \eqno(4.11)$$

\noindent
with $x_1, x_2$ been arbitrary odd Grassmann numbers and $t$
denoting the standard transposition.
The condition (4.10) translates into the following constraints
over the remaining matrix elements

$$\eta^t = \xi^t E A, \ \ \  a = 1+x_1 x_2, \ \ \ det A = 1-x_1 x_2
\eqno(4.12)$$

\noindent
where $E$ denotes the $2\times 2$ antisymmetric block of
$H$ in Eq. (4.10). We assume $Tr(A) \not= 2$ in order that $a$
and $d$ be coprime polynomials. In the notation of Ref. [11],
the unique irreducible
expression for the superdeterminant is

$$ h = (x^2 + \sigma_1 x + \sigma_2)/(x+\sigma_3)\eqno(4.13)$$

\noindent
where the explicit expressions for $\sigma_i \  (i = 1, 2, 3)$ are
obtained there by applying the modified Euclidean algorithm and
are given in explicit form. Substituting our particular values
for the supermatrix elements  we obtain

$$\sigma_1 = -1-Str M, \ \ \ \sigma_2 = - \sigma_3 = 1, \eqno(4.14)$$

\noindent
in such a way that the null polynomial of minimum degree, given
by the product $(x^2+\sigma_1 x+\sigma_2) (x+\sigma_3)$, is [7]

$$P(x) = x^3-(2+Str M)(x^2-x) - 1. \eqno(4.15)$$

Our next example corresponds to case (2) where $a$ and $d$ are
not coprime polynomials, but nevertheless, they do not have a
common factor.
The corresponding $(1+1)\times (1+1)$
supermatrix is

$$M = \left( \matrix{ \sigma & 0 \cr
0 & 0 \cr}\right), \eqno(4.16)$$

\noindent
where $\sigma$ is an even element of the Grassmann algebra such
that $\bar \sigma = 0$ and $\sigma^2 = 0$. In this situation our procedure will
produce a
family of null polynomials. Here,  $a = x-\sigma$
and $d = x$ which are not coprime polynomials according to the definition on
Ref. [11], because the ideal generated by $a$ and $d$ is not the
whole ring of even polynomials over the Grassmann algebra. In
particular, it is not possible to find polynomials $P,Q$ such
that $1 = Pa + Qd$. The basic reason for this is the
impossibility of dividing by $\sigma$. Again, we emphasize the
unintuitive fact that even thought $a$ and $d$ are not relative
primes, they do not possess a common factor. The basic
polynomials are

$$\eqalign{ \tilde F &= x(x-\sigma), \ \ \  F = (x-\sigma)^2 = x^2
- 2x\sigma, \cr
\tilde G &= x^2, \ \ \  G = x(x-\sigma)\cr}\eqno(4.17)$$

\noindent
and we need to consider the corresponding factorization
properties. It is obvious, for example, that $\tilde F$ and
$\tilde G$ have $x$ as a common factor. Surprinsigly, this result
can not be obtained by applying the Euclidean algorithm (or the
modified Euclidean algorithm) to $\tilde F$ and $\tilde G$. The
problem is that the first reminder has body zero, so that we
cannot go on to the second step which requires dividing by this remainder.
Besides, the non existence of a unique factorization theorem is
clearly shown here in the identity $x^2 =
(x+z\sigma)(x-z\sigma)$, which $z$ being an arbitrary complex
number. Choosing $z=1$ leads to the conclusion that $\tilde F$
and $\tilde G$ have two common factors of maximum degree which
are $x$ and $(x-\sigma)$. The same happens with $F$ and $G$.
Thus, after each cancellation is made, we are left with four
possible combinations of the reduced ratios

$${\tilde f_i \over \tilde g_i} = {f_j\over g_j}, \ \ i,j =
1,2,\eqno(4.18)$$

\noindent
where

$$\eqalign{\tilde f_1 = x-\alpha, \ \ \ \tilde f_2 = x, \ \ \ f_1
= x-\alpha, \ \ \ f_2 = x-2 \alpha\cr
\tilde g_1 = x , \ \ \ \tilde g_2 = x+\alpha, \ \ \  g_1 = x, \ \
\ g_2 = x-\alpha.\cr} \eqno(4.19)$$

\noindent
For each possibility one can verify that Eq. (4.18) is indeed
correct. According to our discussion on Section 2, the above
factorization properties lead to four possible null polynomials
given by $P_{ij} (x) = \tilde f_i g_j$. They are

$$\eqalign{P_{11} = x^2 - \alpha x, \ \ \  P_{12} = x^2 - 2\alpha x \cr
P_{21} = x^2, \ \ \ P_{22} = x^2 - \alpha x.\cr} \eqno (4.20)$$

\noindent
Since any linear combination of the above polynomials will be
also annihilated by the supermatrix, we finally obtain two basic null
polynomials which are

$$P_1 = x^2 , \ \ \ \ P_2 = \alpha x.\eqno(4.21)$$

Next we consider the general case (3) where $a(x)$ and $d(x)$
are not coprime and have a maximum common divisor $k(x)$. That is
to say we write

$$a(x) = k(x) a_1 (x), \ \ \  d(x) = k(x) d_1(x), \eqno(4.22)$$

\noindent
where $a_1(x)$ and $d_1(x)$ are coprime polynomials.
In the first place we assume also that $k$ and $a_1$ together
with $k$ and $d_1$ are coprime polynomials
respectively. Each of the polynomials that we have
introduced is monic. This constitutes an extension of the
discussion in Ref. [11] and the next step is to consider the new
factorization properties of the polynomials $\tilde F, \tilde G,
F, G$. We begin by writting

$$\eqalign{\tilde F = (k^A d_1^{A-1} a_1 + Y), \ \ \ \tilde G = k^A
d_1^A ,\cr
F = k^B a_1^B, \ \ \ \ G = (k^B a_1^{B-1} d_1 + Z),\cr}\eqno(4.23)$$

\noindent
where $Y$ and $Z$ are polynomials with body zero and $A=p+1,
B=q+1$. Using the factorization lemma (3.6) of Ref. [11] with
respect to each of the prime polynomials involved we can factor
$\tilde F$ and $G$ in the following way

$$\eqalign{\tilde F &= (k^A + Y_1) (d_1^{A-1} + Y_2) (a_1 + Y_3),\cr
G &= (k^B + Z_1) (a_1^{B-1} + Z_2) (d_1 + Z_3),\cr}\eqno(4.24)$$

\noindent
where $Y_i, Z_i, i= 1,2,3$ are body-zero polynomials. Using the
lemma 3.4 of Ref. [11] together with the fact that $k$, $d_1$, and
$a_1$ are coprime in pairs, we can show that the above
factorization is unique. The condition $\tilde F G = F\tilde G$ and
another use of lemma 3.4 leads to the following identities

$$k^{A+B} = (k^A + Y_1) (k^B + Z_1), \eqno(4.25a)$$
$$d_1^A = (d_1^{A-1} + Y_2) (d_1 + Z_3), \eqno(4.25b)$$
$$a_1^B = (a_1^{B-1} + Z_2) (a_1 + Y_3). \eqno(4.25c)$$

In this way we can identify the basic factorizations
(2.11) by writting

$$\eqalign{R &= (d_1^{A-1} + Y_2), \ \ \ \ S= (a_1^{B-1} + Z_2), \cr
\tilde f &= (k^A + Y_1) (a_1 + Y_3), \ \ \ \   f= k^B (a_1 + Y_3),\cr
\tilde g &= k^A (d_1 + Z_3), \ \ \ \ g = (k^B + Z_1) (d_1 + Z_3).\cr}
\eqno(4.26)$$

Let us observe that the relation $\tilde f g = f\tilde g$ is
immediately realized by virtue of the Eq. (4.25a). Following the
general procedure and modulo accidental cancellations that could
occur in $(k^A + Y_1)/k^A$ or $k^B/(k^B + Z_1)$ for example,  we
identify

$$P(x) = k^{A+B} (a_1 +  Y_3) (d_1 + Z_3) \eqno(4.27)$$

\noindent
as the null polynomial of minimum degree in this case.

A simple specific example of the above case is provided by the
following $(2+2) \times (2+2)$ supermatrix

$$M = \left( \matrix{ 0 & 0 & 0 & \alpha_1 \cr
0 & 1 & \alpha_2 & 0 \cr
0 & \alpha_1 & -1 & 0\cr
\alpha_2 & 0 & 0 & 0\cr}\right), \eqno(4.28)$$

\noindent
where $\alpha_1, \alpha_2$ are odd Grassmann numbers and we define
$\sigma = \alpha_1 \alpha_2$, such that $\sigma^2 =
\sigma\alpha_1 = \sigma\alpha_2 = 0$. Here $A = B = 3.$ The basic
characteristic polynomials are $a(x) = x(x-1)$ and $d(x) =
x(x+1)$ so that we identify

$$k = x,\ \ \ \  a_1 = x-1,\ \ \ \  d_1 = x+1, \eqno(4.29)$$

\noindent
which are indeed coprime in pairs. The basic polynomials are

$$\eqalign{\tilde F &= x^3 (x+1)^2 (x-1) + \sigma x(x+1), \ \ \ \ F=x^3(x-1)^3,
\cr
\tilde G &= x^3 (x+1)^3, \ \ \ \ G= x^3(x-1)^2 (x+1)- \sigma x
(x-1).\cr} \eqno(4.30)$$

The induced factorization properties (4.25) are

$$\eqalign{ x^6 &= (x^3 - \sigma x)(x^3 +\sigma x), \cr
(x+1)^3 &= \left( (x+1)^2 - {\sigma\over 2} (x+1)\right) \left(
x+1+{\sigma \over 2}\right), \cr
(x-1)^3 &= \left( (x-1)^2 + {\sigma\over 2} (x-1)\right) \left(
x-1 - {\sigma\over 2}\right), \cr} \eqno(4.31)$$

\noindent
from where we can read off the values for $Y_i, Z_i$. This,
together with (4.24) and (4.30), allows  to verify the
factorization properties of $\tilde F, \tilde G, F, G$. In this
particular example we have accidental cancellations in such a way
that

$$R = x(x+1) (x+1-\sigma/2), \ \ \ \ S= x(x-1)(x-1+\sigma/2), \eqno(4.32)$$

\noindent
each of which differs from the corresponding expression in Eq.
(4.26) by an extra factor of $x$. The null polynomial of minimum
degree is then

$$P(x) = x^6 + \sigma x^5 - x^4, \eqno(4.33)$$

\noindent
which is of degree six instead of eight, due to the above
mentioned accidental cancellations.

We can further extend the previous case (4.22) by writting the
maximum common divisor $k(x)$ as

$$k(x) = k_0 k_a k_d, \eqno(4.34)$$

\noindent
which explicitly displays the further factorization properties
involved according to the following procedure. Once $k$ has been
identified, it is further written as the product of its minimun degree
factors, which are subsequently rearranged according to the
following convention: those having a common factor
with $a_1 (d_1)$ but not with $d_1 (a_1)$ are called $k_a (k_d)$
respectively while all the remaining factors are included in
$k_0$.  We further demand that  ${k}_{0}$, ${k}_{a}{a}_{1}$ and $
{k}_{d}{d}_{1}$ be coprime in pairs. The previous case corresponded to $k_a =
k_d = 1$. Again,
we start from the expressions (4.23) where $k$ is substituted by
Eq. (4.34) and we look for the factorization properties analagous to Eqs.
(4.24). Here one must be careful enough in keeping together any product
of powers of $k_a$ and $a_1$ or $k_d$ and $d_1$, because the
members of each pair are not respectively coprime. In this way we
obtain

$$\eqalign{ \tilde F &= (k_0^A + \bar Y_1)(k_d^A d_1^{A-1} + \bar
Y_2) (k_a^A a_1 + \bar Y_3)\cr
G &= (k_0^B + \bar Z_1) (k_a^B a_1^{B-1} + \bar Z_2)(k_d^B d_1 +
\bar Z_3)\cr}\eqno(4.35)$$

\noindent
where $\bar Y_i, \bar Z_i, i = 1, 2, 3$ are body zero polynomials.
Using the identity $\tilde F G = F\tilde G$ together with Lemma
(3.4) we extend the factorization properties (4.25) to

$$k_0^{A+B} = (k_0^A + \bar Y_1)(k_0^B+\bar Z_1), \eqno(4.36a)$$

$$k_d^{A+B} d_1^A = (k_d^A d_1^{A-1} + \bar Y_2)(k_d^B d_1 +
\bar Z_3), \eqno(4.36b)$$

$$k_a^{A+B} a_1^B = (k_a^B a_1^{B-1} + \bar Z_2)(k_a^A a_1 + \bar
Y_3).\eqno(4.36c)$$

This time we are not able to directly write $\tilde F, \tilde G,
F, G$ in the way prescribed by Eqs. [2.11]. Instead we
can only arrive to the following general expressions

$$\eqalign{\tilde F = T \tilde f_1 , \  \ \ \tilde G = T \tilde g_1
k_a^A/k_d^B\cr
F = Uf_1 k_d^B/k_a^A, \ \ \ G = Ug_1, \ \ \ \tilde f_1 g_1 = f_1
\tilde g_1, \cr}\eqno(4.37)$$

\noindent
where

$$\eqalign{T = (k_d^A d_1^{A-1} + \bar Y_2), \ \ \  U = (k_a^B
a_1^{B-1} + \bar Z_2), \cr
\tilde f_1 = (k_0^A + \bar Y_1)(k_a^A a_1 + \bar Y_3), \ \ f_1 =
k_0^B (k_a^A \tilde a_1 + \bar Y_3),\cr
\tilde g_1 = k_0^A (k_d^B d_1 + \bar Z_3), \ \ \ g_1 = (k_0^B +
\bar Z_1)(k_d^B d_1 + \bar Z_3).\cr} \eqno(4.38)$$

The Eqs. (4.38) are the generalizations of Eqs. (4.26) and
we verify that $\tilde f_1 g_1 = f_1 \tilde g_1$ is satisfied in
virtue of Eq. (4.36a). Two remarks are now in order: (i) the form
 (4.37) of writting $F$ and $\tilde G$ is rather unpleasant because it
does not clearly exhibits the polynomial character of these
functions. Nevertheless we know that the products $T\ \tilde g_1$
and $U{f_1}$ are in fact divided by $k_d^B$ and $k_a^A$  respectively according
to the factorization equations (4.36b) and (4.36c) .
(ii) the fact that we are not able to write $\tilde F, \tilde G,
F, G$ in the form of Eqs. [2.11] means only that the
method employed does not allow the general identification of a
maximum common divisor in each case, as it happened previously.
Nevertheless, the form (4.37) for the basic polynomials can
also be  used to construct a null polynomial according to the ideas
of Section 3. The definition of the null polynomial in this case
is

$$P(x) = k_a k_d \tilde f_1 g_1 \eqno(4.39)$$

\noindent
and the proof that $P(M)=0$ follows exactly the same steps as
the general discussion of Section 3, with the only difference
that the matrix $N(x) = P(x) (xI-M)^{-1}$ is constructed with the
above $P(x)$. Let us remind the reader that we only need to prove
that $N(x)$ is a polynomial supermatrix. Let us consider the 11
block of the supermatrix $N$. Using (3.13) together with (4.39)
and the expression for $\tilde F$ in (4.37) we can write

$$N_{ij} = -k_a k_d g_1 {\partial \tilde f_1 \over \partial
A_{ji}} - {k_a k_d g_1 \tilde f_1 \over T} {\partial T\over
\partial A_{ji}}. \eqno(4.40)$$

\noindent
The first term of the RHS is clearly of polynomial character. In
order to transform the second term we make use of the property

$${\partial ln \tilde G\over \partial A_{ji}} = 0 = {\partial
lnT\over \partial A_{ji}} + {\partial ln\tilde g_1\over \partial
A_{ji}} + (p+1) {\partial ln k_a\over \partial A_{ji}} - (q+1)
{\partial ln k_d \over \partial A_{ji}}, \eqno(4.41)$$

\noindent
which follows from the factorization (4.37) of $\tilde G$, together
with the fact that $\tilde G$ is just a function of
$D_{\alpha\beta}$, according to Eq. (2.7a). In this way and using
the last relation (4.37) we obtain

$$N_{ij} = k_a k_d f{\partial \tilde g_1 \over \partial A_{ji}} -
k_a k_d g_1 {\partial \tilde f_1 \over \partial A_{ji}} + (p+1)
g_1 \tilde g_1 k_d {\partial k_a\over \partial A_{ji}} - (q+1)
g_1 \tilde f_1 k_a {\partial k_d \over \partial A_{ji}},
\eqno(4.42)$$

\noindent
which leads to the conclusion that the block-matix $N_{ij}$ is
indeed polynomial.

The proof for the remaining blocks of $N(x)$ follows exactly
along the arguments of Section 3, except that now more terms are
involved in complete analogy with Eq. (4.42). The final
conclusion is that $N(x)$ is indeed polynomial thus leading to
$P(M) = 0$.

Finally we present an specific example of the previous case. Let
us consider the $(2+2) \times (2+2)$ supermatrix

$$M = \left( \matrix{ 1 & 0 & 0 & \alpha _ 1 \cr
0 & 0 & \alpha_2 & 0\cr
0 & \alpha _1 & 0 & 0 \cr
\alpha_2 & 0 & 0 & 0 \cr}\right) \eqno(4.43)$$

\noindent
with $\sigma = \alpha_1 \alpha_2$ as previously introduced. Now
we have $a(x) = x(x-1)$ and $d(x) = x^2$, with $A=B=3.$ The
maximum common divisor is $k = x$ and our conventions to denote
the  factors of $k$ leads to

$$k_0 =k_a = 1, \ \ \ \ k_d = x, \ \ \ \ \  a_1 = x-1, \ \ \ \ d_1 =
x.\eqno(4.44)$$

The basic polynomials are

$$\eqalign{\tilde F = x^5(x-1) - \sigma x^3, \ \ \ \ F = x^3
(x-1)^3, \cr
\tilde G = x^6,  \ \ \ \  G = x^4 (x-1)^2 + \sigma x^2
(x-1).\cr}\eqno(4.45)$$

In this case, the factorization (4.36a) does not occur and the
remaining ones are

$$\eqalign{x^9 = (x^5 + \sigma x^3 (x+1))(x^4 - \sigma x^2 (x+1))\cr
(x-1)^3 = ((x-1)^2 + \sigma (x-1))(x-1- \sigma) \cr}\eqno(4.46)$$

\noindent
which correspond to (4.36b) and (4.36c) respectively. From these
expressions we can read off the polynomials $\bar Y_2, \bar Y_3,
\bar Z_2, \bar Z_3$ and verify the factorizations (4.35) for
$\tilde F$ and $G$ with the understanding that $(k_0^A + \bar
Y_1)$ and $(k_0^B + \bar Z_1)$ should be replaced by one. Going
back to the Eqs. (4.38) we find

$$\eqalign{ T = x^3 (x^2 + \sigma (x+1)), \ \  U = (x-1)(x-1+
\sigma)\cr
\tilde f_1 = x-1-\sigma, \ \  f_1 = x-1-\sigma \cr
\tilde g_1 = x^2 (x^2 - \sigma(x+1)), \ \  g_1 = x^2 (x^2-
\sigma(x+1)\cr} \eqno(4.47)$$

\noindent
and the null polynomial (4.39) is given by

$$P(x) = x^6 - x^5 (1+3\sigma) + \sigma x^3.\eqno(4.48)$$

It is interesting to observe that accidental cancellations which occur
in this case, that we are not able to describe in general,
allow us to rewrite the expressions (4.37) exactly in the form
(2.11), (2.12) with the following identifications

$$\eqalign{ R = x^2 (x^2 + \sigma (x+1)), \ \ \ \  S = x^2
(x-1)(x-1+\sigma),\cr
\tilde f = f = x(x-1-\sigma), \ \ \ \  \tilde g = g = x^2 - \sigma
(x+1).\cr} \eqno(4.49)$$

In this way we can find another null polynomial of  degree lower
than (4.42), which is given by

$$P_1(x) = fg = x^4 - (1+3\sigma) x^3 + \sigma x. \eqno(4.50)$$

The simple form of the supermatrix (4.43) permits an easy
verification that $P_1(M) = 0$.
\vskip 1.4pc
\noindent
5.- SUMMARY

Given an arbitrary supermatrix $M$ and starting from $ Sdet\left( xI-M
\right)$, which is naturally written as a ratio of  polynomials, we have given
a prescription to construct various types of null polynomials. We have also
proved that each of them is annihilated by $M$, thus providing an extension of
the Cayley-Hamilton theorem for supermatrices. At the level of some particular
cases we have also extended  some results of Ref. [11] by giving  a
constructive procedure to produce the required factorizations needed to
construct what we have called null polynomials of minimum degree,  for the case
where $a(x)$ and $d(x)$ have a common factor.

In order to put our work in the right perspective and to suggest some possible
lines of further research, we now make a few comments. We have called
"characteristic" the polynomial ${\cal P}( x )$  defined in Eq.(2.10) because
it is the one that can be directly associated with an arbitrary supermatrix,
independently of the factorization properties of the numerator and denominator
of $ Sdet\left( xI-M \right)$. Nevertheless, this polynomial carries very
little information regarding the odd blocks of $M$ and  so far we have not
studied to what extent it really characterizes the supermatrix. Our guess is
that the null polynomials of minimum degree, defined in Eq.(2.13) and
emphasized in the examples,  will be much more effective in this respect.
Nevertheless, we are still lacking a completely general procedure or
classification to determine when there would exist a maximum common divisor of
the polynomials $ \tilde{F}$ and $ \tilde{G}$ \ \   \ ( $F$ and $G$) \  that
are the b!
uilding blocks of $ Sdet\left( xI
using the euclidean algorithm. From our point of view,  it is clear that these
matters require further understanding.

\

The work of both authors has been partially supported by the
grant DGAPA-UNAM-IN100691. LFU also acknowlegdes
support from the project  CONACyT-0758-E9109. Both
authors thank  Dr. Horacio Tapia for innumerable
discussions and clarifications regarding some mathematical aspects of the work
presented here.

\vfill\eject
\noindent
APPENDIX A

Here we give an explicit proof of the equivalence between the two
ways of calculating $S det(M)$ which were written in Eq.
(2.5). We are assuming that $det (A) \not= 0$ and $det(D) \not= 0$.

Let us consider

$$\Omega = ln \left[ {det(A-BD^{-1} C)\over det(A)} \right], \eqno(A.1)$$

\noindent
which we can rewrite as

$$\Omega = ln \ det (I-A^{-1} B D^{-1} C) = Tr \ ln (I-A^{-1} B D^{-1} C).
\eqno(A.2)$$

\noindent
The expansion of $ln(1-z)$ in power series leads to

$$\Omega = - \sum^\infty_{n=1} {1\over n} Tr (A^{-1}B D^{-1}
C)^n.\eqno(A.3)$$

\noindent
Notice that the series expansion indeed stops at some power $N$
given by the degree of nilpotency of $A^{-1}BD^{-1}C$. Now we use
the cyclic property of the trace together with the fact that
$B_{i\alpha}$ and $C_{\beta i}$ are odd Grassmann numbers, to
move $A^{-1}$ and $B$ from the beginning to the end of each term
in the series expansion (A.3). In this way we obtain

$$\Omega = + \sum^\infty_{n=1} {1\over n} Tr(D^{-1} C A^{-1}
B)^n,\eqno(A.3)$$

\noindent
where the change of sign occurs because $B$ has to be moved
through an odd number of anticommuting Grassmann matrices. By
reversing the previous  expansion argument we conclude that

$$\Omega = - Tr \ ln (I-D^{-1} CA^{-1}B) = - ln \left[ {det (D -
CA^{-1}B)\over det (D)} \right]. \eqno(A.4)$$

\noindent
Thus we have shown that

$${det (A-B D^{-1}C)\over det (A)} = {det (D)\over det (D-CA^{-1}B)},
\eqno(A.4)$$

\noindent
which readily implies the Eq. (2.5).

\vskip 1.4pc
\noindent
APPENDIX B

In this appendix we collect some basic results of Ref. [11]  which
we have used in this work. As far
 as possible we follow the
notation and conventions of this reference and also we use their
numeration for the respective lemmas and corollaries.

We consider a Grassmann algebra $\Lambda$ over the complex numbers
$C\hskip-7pt I \ $. Any element $a \in \Lambda$ is a sum of the
body $\bar a \in C \hskip-7pt I \ $ plus the nilpotent element
$s(a)$ called the soul. The ring of polynomials over this
Grassmann algebra is denoted by $\Lambda_0[x]$ and consists of all
polynomials

$$f(x) = a_0 x^n + a_1 x^{n-=1} + \cdots + a_n \eqno(B.1)$$

\noindent
where $a_k$ are even elements of the Grassmann algebra. The set of
nilpotent elements of $\Lambda_0[x]$ is denoted by ${\cal Q} =
s(\Lambda_0) [x]$.

Two polynomials $S$ and $T$ in $\Lambda_0[x]$ are coprime if the
ideal generated by  $S$ and $T$ is the whole ring $\Lambda_0[x]$.

Lemma 3.4: Let $S, T, S_1, T_1$ in $\Lambda_0 [x]$ and suppose
$\bar S = \bar S_1, \bar T = \bar T_1$ with $S$ and $T$
being coprime. If $ST = S_1 T_1$ then $S_1 = (1+R)S$ and $T_1 =
(1/(1+R))T$ with $R \in$ ${\cal P}$. If moreover $S$ and $S_1$ are monic,
then $S = S_1$ and $T = T_1$.

Corollary 3.8: Let $S$ and $T$ be coprime monic polynomials in
$\Lambda_0[x]$ and let $R$ be in ${\cal Q}$ such that $deg(R) < deg
(ST)$. Then there exist $R_1$ and $R_2$ in ${\cal Q}$ such that $ST+R=
(S+R_1) (T+R_2)$, with $deg(R_1) < deg (S)$ and $deg(R_2) < deg
(T)$.

\vfill\eject
\noindent
REFERENCES

\

\item{[1]} For a review see for example Birmingham D, Blau M,
Rakowski M and Thompson G 1991 {\it Phys. Rep.} {\tit 209} 129

\item{[2]} Jones V 1985 {\it Bull AMS} {\tit 12} 103 and 1989
{\it Pacific J. Math.} {\tit 137} 312

\item{ \ } Freyd P, Yetter D, Hoste J, Lickorish W,
Millet K and Ocneanu A 1985 {\it Bull AMS} {\tit 12} 239

\item{ \ } Kauffman L 1987 {\it Topology} {\tit 26} 395

\item{ \ }Witten E 1989 {\it Commun. Math. Phys.}
{\tit 121} 351
and 1989 {\it Nucl. Phys.} {\tit B322} 629

\item{[3]} Witten E 1988/89 {\it Nucl. Phys.} {\tit B311} 46

\item{[4]} Horne J H 1990 {\it Nucl. Phys.} {\tit B334} 669

\item{[5]} Nelson J E, Regge T and Zertuche F 1990 {\it Nucl.
Phys.} {\tit B339} 516

\item{[6]} Koehler K, Mansouri F, Vaz C and Witten L 1990 {\it
Mod. Phys. Lett.} {\tit A5} 935; 1990 {\it Nucl. Phys.} {\tit
B341} 167 and 1991 {\it Nucl. Phys.} {\tit B348} 373

\item{[7]} Urrutia L F, Waelbroeck H and Zertuche F 1992 {\it Mod.
Phys. Lett.} {\tit A7} 2715

\item{[8]}  Mandelstam S 1979 {\it Phys. Rev.} {\tit D} 2391

\item{ \ } Loll R 1992 "Loop Approaches to Gauge Field Theory",
Syracuse preprint SU-GP-92/6-2

\item{[9]} Urrutia L F and Morales N 1992  "The Cayley-Hamilton
theorem for supermatrices", to be published in {\it J. Phys.} {\tit A :}
{\it Math.Gen.}, letters section.

\item{[10]} See for example De Witt B 1984 {\it Supermanifolds}
(Cambridge University Press, Cambridge)

\item{[11]} Kobayashi Y and Nagamachi S 1990 {\it J. Math. Phys.}
{\tit 31} 2726

\item{[12]} See for example Nering E D 1970 {\it Linear Algebra and
Matrix Theory} (Second edition, John Wiley, New York)

\item{[13]} Urrutia L F, Waelbroeck H and Zertuche F, preprint
ICN-UNAM 1992, {\it The Algebra of Supertraces for 2+1 Super de
Sitter Gravity}, to be published in the Proceedings of the
Workshop on Harmonic Oscillators, Maryland, March 1992

\end